\documentclass[prb,twocolumn,floatfix,amsmath,amssymb]{revtex4-1}
\usepackage{times}
\pdfoutput=1
\usepackage{bbold}
\usepackage{graphicx}
\usepackage{dcolumn}
\usepackage{bm}
\usepackage{color}
\usepackage{hyperref}
\renewcommand{\vec}{\mathbf}

\newcommand{\mi}{\mathrm{i}}


\begin{document}

\title{Quantum oscillations in $\mathrm{YBa_{2}Cu_{3}O_{6+\delta}}$ from  an incommensurate $d$-density wave order}

\author{Jonghyoun Eun}
\author{Zhiqiang Wang}
\author{Sudip Chakravarty}
\affiliation{Department of Physics and Astronomy, University of
California Los Angeles, Los Angeles, California 90095-1547, USA}

\date{\today}

\begin{abstract}
We consider quantum oscillation  experiments in $\mathrm{YBa_{2}Cu_{3}O_{6+\delta}}$ from the perspective of an incommensurate  Fermi
surface reconstruction using an exact transfer matrix method and the Pichard-Landauer formula for the conductivity. The specific density wave order considered is a  period-8 $d$-density wave in which
the  current density is unidirectionally modulated. The current modulation is also naturally accompanied by a  period-4 site charge modulation in the same direction, which
 is consistent with recent magnetic resonance measurements. In principle Landau theory also allows for a period-4 bond charge modulation, which is not discussed, but should be simple to incorporate in the future.  This scenario  leads to a natural, but not a unique,  explanation of why only oscillations from
 a single electron pocket is observed, and a hole pocket of roughly twice the frequency as dictated by two-fold commensurate order, and the corresponding Luttinger sum rule,  is not observed. However, it is possible that even
 higher magnetic fields will reveal a hole pocket of  half the frequency of the electron pocket or smaller. This may be at the borderline of achievable high field measurements because at least a few  complete 
 oscillations have to be clearly resolved.
\end{abstract}l

\maketitle

\section{Introduction}
The discovery of  quantum oscillations~\cite{Doiron-Leyraud:2007} in the Hall coefficient ($\mathrm{R_{H}}$) of   hole-doped high temperature superconductor  $\mathrm{YBa_{2}Cu_{3}O_{6+\delta}}$ (YBCO)  in high magnetic  fields approximately between
$ 35 - 62\; \rm{T}$ was an important event.~\cite{Chakravarty:2008}  Although the original measurements were performed in the underdoped regime, close to 10\% hole doping,  later measurements have also revealed clear oscillations for $\mathrm{YBa_{2}Cu_{4}O_{8}}$ (Y248), which corresponds to about 14\% doping.~\cite{Yelland:2008,Bangura:2008}  Fermi surface reconstruction due to  a density wave order that could
arise if superconductivity is ``effectively destroyed'' by high magnetic fields has been a promising focus of  attention.~\cite{Millis:2007,Chak2:2008,Dimov:2008,Podolsky:2008,Chen:2009,Yao:2011}
Similar quantum oscillations in the $c$-axis resistivity in  $\mathrm{Nd_{2-x}Ce_{x}CuO_{4}}$ (NCCO)~\cite{Helm:2009,*Helm:2010,*Kartsovnik:2011} have been easier to interpret in terms of a two-fold commensurate density wave order, even quantitatively, including
magnetic breakdown effects.~\cite{Eun:2010,*Eun:2011}

At this time, in YBCO, there appears to be no general agreement about the precise nature of the translational symmetry breaking.  A pioneering idea invoked an order corresponding to period-8  anti-phase spin stripes.~\cite{Millis:2007} The emphasis there was to show how over a reasonable range of parameters the dominant Fermi pockets are electron pockets, thus explaining the observed negative Hall coefficient.
 At around the same time one of us suggested a two-fold commensurate $d$-density wave (DDW) order that could also explain the observations.~\cite{Chak2:2008}There are several reasons for such a choice.
 One of them  is that the presence of both hole and electron pockets with differing scattering rates leads to a natural explanation~\cite{Chak2:2008} of oscillations of $\mathrm{R_{H}}$. An incommensurate period-8 DDW was also considered.~\cite{Dimov:2008} Fermi surfaces resulting from this order are very similar to those due to  spin stripes. The lack of Luttinger sum rule and a multitude  of possible   reconstructed Fermi surfaces appeared to have  little constraining power in a Hartree-Fock mean field theory. However, since then many experiments that indicate the importance of stripe physics~\cite{Taillefer:2009} and even possible unidirectional charge order have led us to reconsider the period-8  DDW.

We enumerate below further motivation for this reconsideration.
\begin{itemize}
\item Tilted field measurements have revealed  spin zeros in quantum oscillations, which indicate that the symmetry breaking order parameter is a singlet instead of a triplet.~\cite{Ramshaw:2011,Sebastian:2011}
 The chosen order parameter is therefore likely to be a singlet  particle-hole condensate rather than a triplet. The nuclear magnetic resonance (NMR) in high fields indicate that there is no spin order but  a period-4  charge order that develops at low temperatures.~\cite{Julien:2011}
 \item As long as the CuO-plane is square planar, the currents induced by the DDW cannot induce net magnetic moment to couple to the nuclei in a NMR measurement.  Any deviation from the square planar character could give rise to a NMR signal.~\cite{lederer:2011} To the extent these deviations are small the effects will be also small. Thus the order is very effectively hidden.~\cite{Nayak:2000,Chakravarty:2001} 
\item While commensurate models can explain measurements in NCCO, it appears to fail to explain the measurements in YBCO. Luttinger sum rule leads to a concommitatnt hole pocket with an oscillation frequency roughly about twice the frequency of the electron pocket ($\sim 500 \; \rm{T}$). Despite motivated search no such frequency has been detected. In contrast, for  incommensurate period-8 DDW the hole pockets can be quite {\em small} for a range of  parameters. In order to  convincingly detect such an oscillation, it is necessary to perform experiments in much higher fields than currently practiced. This may be a resolution of the non-obervation of the hole pocket.  However,  a tantalizing evidence of  a frequency ($250\; \rm{T}$)  has been reported in a recent $85 T$ measurement,~\cite{Singleton:2010} but its confirmation will require further experiments.
 \end{itemize}

In Sec. II we describe our model, while in Sec. III we outline the transfer matrix calculation of the conductivity. In Sec. IV, we describe our results, most importantly the quantum oscillation spectra. The final section, V,  contains a discussion and an overall outlook.

\section{The Model}
\subsection{Band structure}
\begin{figure}[htb]
\begin{center}
\includegraphics[width=\linewidth]{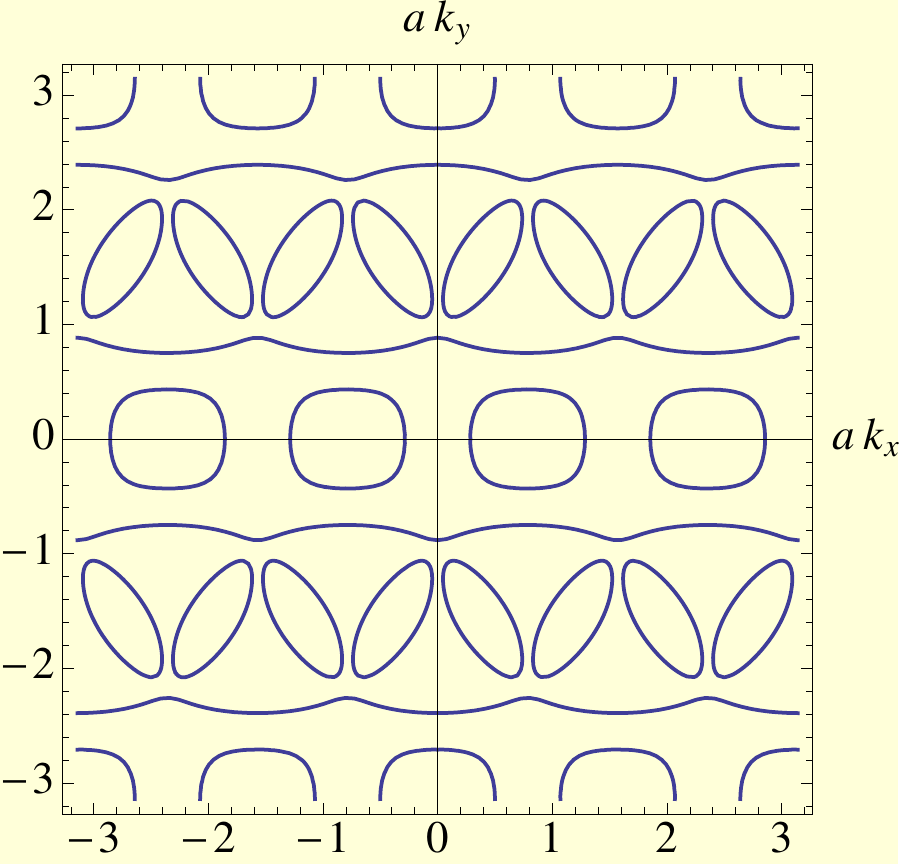}
\caption{Reconstructed Fermi surafces  with ${\bf Q}=\frac{\pi}{a}(\frac{3}{4},1)$,  $W_{0}=0.65 t$ , $V_c=0.05 t$, and  $\mu=-0.83 t$. There are electron pockets, hole pockets and open orbits. The electron pocket frequency corresponds to $530 \text{T}$ and the hole pockets to $280 \text{T}$. The doping corresponds to $12.46\%$. Note that the figure is shown in the extended BZ for clarity.}
\label{fig:highC}
\end{center}
\end{figure}

The parametrization of the single particle band structure for YBCO from angle resolved photoemission spectroscopy (ARPES)  is not entirely straightforward because cleaving at any nominal doping leads to an overdoped surface. Nonetheless an interesting attempt was made to reduce the doping by a potassium overlayer.~\cite{Hossain:2008}  Further complications arise from bilayer splitting and chain bands. Nonetheless, the inferred band structure appears to be  similar to other cuprates where ARPES is a more controlled probe.~\cite{Damascelli:2003}  Here we shall adopt a  dispersion that has become common and has its origin in a local density approximation (LDA) based calculation,~\cite{Pavarini:2001} which is
\begin{equation}
\begin{split}
\epsilon_\vec{k}&= -2t(\cos k_x + \cos k_y)+ 4t'\cos
k_x \cos k_y\\&- 2t''(\cos 2k_x + \cos 2k_y).
\end{split}
\end{equation}
The band parameters are chosen to be
$t~=~0.15 eV$, $t'~=~0.32t$, and $t''~=~0.5t'$.~\cite{Pavarini:2001}  The only difference with the conventional  LDA band structure is a rough renormalization of $t$ (from $0.38 eV$ to $0.15 eV$), which is supported by many ARPES
experiments that find that   LDA overestimates the bandwidth. A more recent ARPES measurement on thin films paints a somewhat more complex picture.~\cite{Sassa:2011}
\begin{widetext}
\subsection{Incommensurate DDW  without disorder}
An Ansatz for period-eight  incommensurate DDW~\cite{Dimov:2008}  involves the wave vector   ${\bf  Q}=(\frac{\pi}{a},\frac{\pi}{a})-\frac{\pi}{a}(2\eta, 0)=\frac{\pi}{a} (\frac{3}{4},1)$ for $\eta=1/8$. 
With the 8-component spinor defined by
$\chi_{\bf k}^{\dagger} = (c_{{\bf k}^{\dagger},\alpha }, c_{{\bf k+Q},\alpha }^{\dagger}, c_{{\bf k}+2{\bf Q},\alpha }^{\dagger}, \ldots c_{{\bf k}+8{\bf Q}}^{\dagger})$, the Hamiltonian without disorder can be written as 
\begin{equation}
{\cal H}=\sum_{{\bf k},\alpha} \chi_{{\bf k}\alpha}^{\dagger} Z_{{\bf k},\alpha}\chi_{{\bf k}\alpha}
\label{eq:highC}
\end {equation} 
The up and down spin sector eigenvalues merely duplicate each other, and we can consider simply one of them:
\begin{equation}
Z_{\bf k}=\left(\begin{array}{cccccccc}\epsilon_{\bf k}-\mu & iG_{\bf k} & V_{c} & 0 & 0 & 0 & V_{c} & -iG_{{\bf k}+7{\bf Q}}  \\\text{c.c} & \epsilon_{\bf k+Q}-\mu & iG_{{\bf k}+{\bf Q}}  & V_{c} & 0 & 0 & 0 & V_{c}\\V_{c} & \text{c.c} & \epsilon_{{\bf k}+2{\bf Q}}-\mu & iG_{{\bf k}+2{\bf Q}} & V_{c} & 0 & 0 & 0 \\0 & V_{c} &  \text{c.c}  & \epsilon_{{\bf k}+3{\bf Q}}-\mu &  iG_{{\bf k}+3{\bf Q}}  &V_{c} & 0 & 0 \\0 & 0 & V_{c} & \text{c.c}  & \epsilon_{{\bf k}+4{\bf Q}}-\mu &  iG_{{\bf k}+4{\bf Q}}&V_{c} & 0 \\0 & 0 & 0 & V_{c} &  \text{c.c} & \epsilon_{{\bf k}+5{\bf Q}}-\mu & iG_{{\bf k}+5{\bf Q}} & V_{c} \\V_{c} & 0 & 0 & 0 & V_{c} & \text{c.c} &  \epsilon_{{\bf k}+6{\bf Q}}-\mu &  iG_{{\bf k}+6{\bf Q}}  \\  iG_{{\bf k}+7{\bf Q}} & V_{c}& 0 & 0 & 0&V_{c}  &  \text{c.c} & \epsilon_{{\bf k}+7{\bf Q}}-\mu \end{array}\right),
\end{equation}
where $G_{\bf k}=(W_{\bf k} - W_{\bf k+Q})/2$, 
and the DDW gap is
$W_\vec{k} = \frac{W_0}{2}(\cos k_x - \cos k_y)$. On symmetry grounds,  one can  quite generally expect that an  incommensurate DDW with wave vector $\bf Q$ will induce a charge density wave (CDW) of wave vector $2{\bf Q}$.~\cite{Nayak:2000} This fact is taken into account by explicitly incorporating a period-4 CDW by introducing the real matrix elements $V_{c}$.  The chemical
potential $\mu$  and the DDW gap amplitude  $W_{0}$ can be adjusted to give the desired quantum oscillation frequency of the electron pocket
as well as the doping level.
The Fermi surfaces corresponding to the spectra of Eq.~(\ref{eq:highC}) (an example is shown in Fig.~\ref{fig:highC}) are  not essentially different from the  mean field theory of $1/8$ magnetic antiphase stripe order.~\cite{Millis:2007} This higher order commensuration  generically produces complicated Fermi surfaces, involving open orbits, hole pockets, and electron pockets. 
\end{widetext}

\begin{figure}[htb]
\begin{center}
\includegraphics[width=\linewidth]{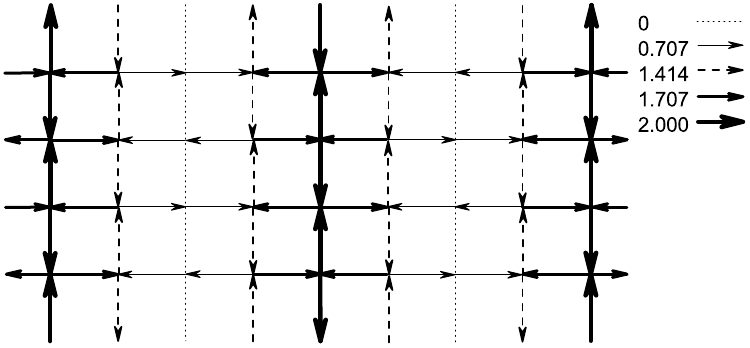}
\caption{Current pattern for ${\bf Q}=(\frac{3\pi}{ 4a}, \frac{\pi}{a})$. The relative magnitudes of the currents are depicted by the arrows in the legend.  Note the antiphase domain wall structure.}
\label{fig:IC-pattern}
\end{center}
\end{figure}

To picture the current modulation and to define the order parameter of period-8 DDW in the real space Hamiltonian we need to calculate $ \langle c_{{\bf R}'}^\dag c_{\bf R}\rangle $ for ${\bf R}'\ne {\bf R}$.   We get,  correcting here a mistake in Ref.~\onlinecite{Dimov:2008},
\begin{equation}
\label{cRR}
\begin{split}
    \langle c_{{\bf R}'}^\dag c_{\bf R}\rangle &= \frac{1}{N}\sum_{{\bf k}'{\bf k}}\langle c_{{\bf k}'}^\dag
   c_{{\bf k}}\rangle \exp\left[-i({\bf k}'\cdot{\bf R}'-{\bf k}\cdot{\bf R})\right]\\
   &=\pm \frac{iW_0}{2}(-1)^{n'+m'}\widetilde{V}_{{\bf R}',{\bf R}},
\end{split}
\end{equation}
where ${\bf R}'=(m'a,n'a)$, and $\widetilde{V}_{{\bf R}',{\bf R}}$ is
\begin{equation}
\begin{split}
    \widetilde{V}_{{\bf R}',{\bf R}}=&\Bigg[\frac{1+\cos 2\pi\eta}{2}(\delta_{{\bf R}',{\bf R}+a{\bf \hat{x}}}+\delta_{{\bf R}',{\bf R}-a{\bf\hat{x}}})\\
       & -(\delta_{{\bf R}',{\bf R}+a{\bf \hat{y}}}+\delta_{{\bf R}',{\bf R}-a{\bf\hat{y}}})\Bigg]\cos 2m'\pi\eta\\
        &+\frac{\sin 2\pi\eta\sin
        2m'\pi\eta}{2}(\delta_{{\bf R}',{\bf R}+a{\bf \hat{x}}}-\delta_{{\bf R}',{\bf R}-a{\bf\hat{x}}}).\\
  \end{split}
\end{equation}
The current pattern is then
\begin{equation}
\label{current}
\begin{split}
    J_{{\bf R}',{\bf R}}&=i[\langle c_{{\bf R}'}^\dag c_{\bf R}\rangle-\langle c_{\bf R}^\dag
        c_{{\bf R}'}\rangle]\\
    &=- W_{0}(-1)^{n'+m'}\widetilde{V}_{{\bf R}',{\bf R}},
\end{split}
\end{equation}
which  is drawn in Fig.~\ref{fig:IC-pattern}.  The  incommensurate $d$-density wave  order parameter is  proportional to
\begin{equation}
\widetilde{W}_{\vec{R'},\vec{R}} = \frac{\mi W_0}{2}(-1)^{n' + m'} \widetilde{V}_{\vec{R'},\vec{R}}.
\end{equation}

\subsection{The real space Hamiltonian including disorder}

In  real space, the Hamiltonian in the presence of both disorder and magnetic field is

\begin{equation}\label{eq:hamiltonian}
\begin{split}
	H=&\sum_{\vec{R}} \left[V(\vec{R})+2V_c \cos(\pi m/2)\right]c_\vec{R}^\dag c_\vec{R}\\
	&+\sum_{\vec{R'},\vec{R}}
		t_{\vec{R'},\vec{R}}~\mathrm{e}^{\mi a_{\vec{R'},\vec{R}}} c^\dag_\vec{R'} c_\vec{R}\\
	&+\sum_{\vec{R'},\vec{R}}
		\widetilde{W}_{\vec{R'},\vec{R}}~\mathrm{e}^{\mi a_{\vec{R'},\vec{R}}} c^\dag_\vec{R'} c_\vec{R}+h.c.
\end{split}
\end{equation}
Here $t_{\vec{R'},\vec{R}}$ defines  the band  structure: the nearest neighbor, the next nearest neighbor, and the third nearest neighbor hopping terms: $t$, $t'$, $t''$.  And $2V_c \cos(\frac{\pi}{2} m)$ is responsible for the period-four charge stripe order, where $m$ is $\vec{R}\cdot \hat{x} / a$ and $a$ is lattice spacing. We include correlated disorder in the form~\cite{Jia:2009}
\begin{equation} 
V(\vec{R}) = \frac{g_V}{2\pi l^2_D} 
	\int ~ d\vec{x} ~ e^{-\frac{|\vec{R}-\vec{x}|^2}{2 l^2_D}} ~ u(\vec{x}),
\end{equation}
where $l_{D}$ is the disorder correlation length and 
 the disorder averages are 
 $\langle u(\vec{x})\rangle=0$ and $\langle
u(\vec{x})u(\vec{y})\rangle=\delta(\vec{x}-\vec{y})$; the disorder intensity
is set by $g_V$. 

 While white noise disorder seems to be more appropriate for
NCCO with intrinsic disorder, correlated disorder may be more relevant to relatively clean YBCO samples in the range of well ordered chain 
compositions. Thus, here we shall focus on correlated disorder.
A constant perpendicular magnetic field $B$ is included via the Peierls phase factor 
$a_{\vec{R'},\vec{R}}=\frac{e}{\hbar c}\int_\vec{R}^\vec{R'}\vec{A}\cdot\mathrm{d}\vec{l}$, 
where $\vec{A}=(0,-Bx,0)$ is the vector potential in the Landau gauge;
 the lattice vector ${\bf R}'=(m'a,n'a)$ is defined by an  arbitrary set of integers.

\section{The transfer matrix method}
The transfer matrix technique is a powerful method to compute conductance oscillations. It requires 
neither quasiclassical approximation nor {\em ad hoc} broadening of the
Landau level to incorporate the effect of disorder. Various models of disorder, both long and short-ranged, can be studied {\em ab initio}. The mean field Hamiltonian, being a quadratic non-interacting Hamiltonian, leads to a Schr\"odinger equation for the site amplitudes, which is then recast in the form of a transfer matrix; the  derivation has been discussed in detail previously.~\cite{Jia:2009,Eun:2010} The conductance is then calculated by a formula that is well known in the area of mesoscopic physics, the Pichard-Landauer formula.~\cite{Pichard:1986,*Fisher:1981} This yields Shubnikov-de Haas oscillations of the $ab$-plane resistivity, $\rho_{ab}$. 

We consider a quasi-1D system, $N\gg M$, with a periodic boundary
condition along y-direction. Here $Na$ is the length in the $x$-direction and $Ma$ is the length in the $y$-direction.  Let $\Psi_n = (\psi_{n,1},\psi_{n,2}, \ldots, \psi_{n,M})^T$, $n=1, \dots N$,
be the amplitudes on the slice $n$ for an
eigenstate with a given energy. Then the amplitudes between the 
successive slices depending on the Hamiltonian  must form a given transfer matrix, $\mathbb{T}$.

The complete set of Lyapunov exponents, $\gamma_{i}$, of 
$\lim_{N\to\infty}({\cal T}_{N}{\cal T}_{N}^{\dagger})$, where
${\cal T}_{N}=\prod_{j=1}^{j=N}{\mathbb T}_{j}$ determine   
 the conductance, $\sigma_{ab}(B)$ from the Pichard-Landauer formula:
\begin{equation}
\sigma_{ab}(B) = \frac{e^{2}}{h} \text{Tr}\sum_{j=1}^{2M}\frac{2}{({\cal
T}_{N}{\cal T}_{N}^{\dagger})+({\cal T}_{N}{\cal
T}_{N}^{\dagger})^{-1}+2}.
\end{equation}
In this work we have chosen $M=30$  and $N$ of the order of $10^{5}$. This guaranteed $4\%$ accuracy of the smallest Lyapunov exponent. Note that at each step we have to invert a $4M\times 4M$ matrix and numerical errors prohibit much larger values of $M$.
\section{Results}
\subsection{Specific heat without disorder}
 The coefficient of the linear specific heat is 
\begin{equation}
\gamma=\frac{\pi^{2}}{3} k_{B}^{2}\; \rho(0) .
\end{equation}
The density of states $\rho(\omega)$ measured with respect to the Fermi energy can be easily computed by taking into account all eight bands in the
irreducible part of the Full Brillouin zone and  a factor of 2 for spin. A Lorentzian broadening of the $\delta$-functions was used in computing the density of states.
Although this is useful for numerical computation, the smoothing is a rough way of incorporating the effect of disorder on the density of states.

\begin{figure}[htbp]
\begin{center}
\includegraphics[scale=.85]{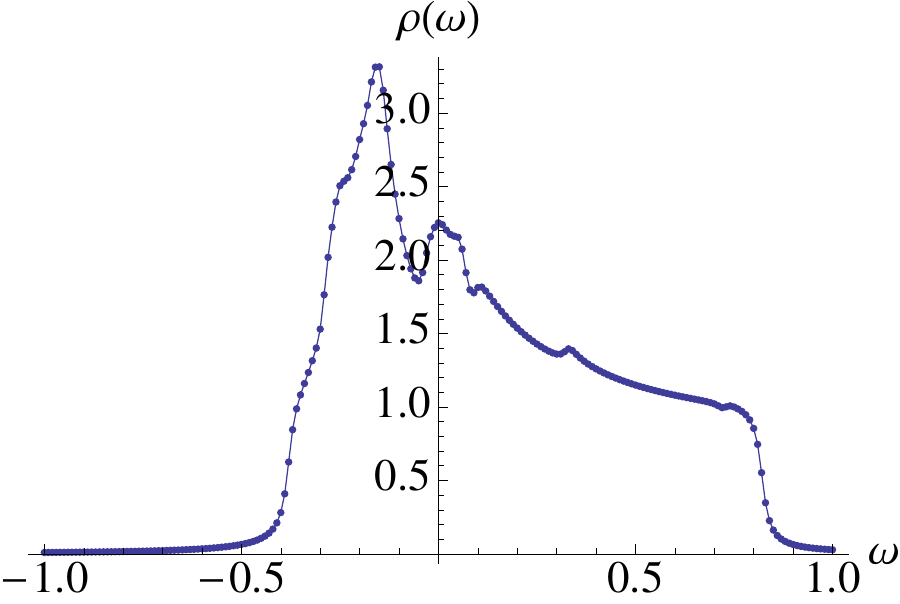}
\caption{Total density of states, $t=0.15$, including eight bands in the reduced Brillouin zone per layer. The horizontal axis is in terms of electron volts and the vertical axis is a pure number, that is, the number of states. The rounding at the tails is due to the Lorentzian broadening of the $\delta$-functions by $\Gamma=0.1 t$. The remaining parameters are the same as in Fig.~\ref{fig:highC}. Further smoothing will reduce the density of states at the Fermi energy and lower the value of $\gamma$.} 
\label{Fig1}
\end{center}
\end{figure}

For a single CuO-layer we get,  
\begin{equation}
\gamma\approx  5.4 \frac{mJ}{mole.K^{2}} ,
\end{equation}
where we have used the density of states at the Fermi energy from numerical calculation to be approximately 2.3 states/eV, as shown in the Figure~\ref{Fig1}. Including both layers $\gamma=2\times 5.4=10.8 \frac{mJ}{mole. K^{2}}$, approximately a factor of 2 larger  than the observed $ 5\frac {mJ}{mole. K^{2}}$ at $45 T$.~\cite{Riggs:2011}

\subsection{Charge modulation without disorder}
Before we carry out an explicit calculation it is useful to make a qualitative estimate. For $V_{c}=0.05 t$, the total charge gap is  
to $4V_{c}=0.2 t$. In order to convert to modulation of the charge order parameter, we have to divide by a suitable  coupling constant. In 
high temperature superconductors, all important coupling constants are of the order  bandwidth, which is $8 t= 1.2 eV$. Taking this as 
a rough estimate, we deduce that the charge modulation is $0.025 e$, expressed in terms of electronic charge.

To explicitly calculate  charge modulation at a site, we diagonalize the $8\times 8$  Hamiltonian matrix $Z(k_x, k_y)$
for each {\bf k} in the reduced Brillouin zone (RBZ). We get 8 eigenvalues
and the corresponding eigenvectors: $E_{n,k_x,k_y}$  and $\psi_{n,k_x,k_y}$ for $n=1,2,...,8$.
The eigenvector $\psi_{n,k_x,k_y}$ has  eight components of the  form:
\begin{equation}
\psi_{n,k_x,k_y}=(\alpha_{n,k_x,k_y}(1), \alpha_{n,k_x,k_y}(2),....,\alpha_{n,k_x,k_y}(8))
\end{equation}
Then the  wave function in the real space for each state $\{n,k_x,k_y\}$ is
\begin{equation}
\psi_{n,k_x,k_y}(\vec{R})=\sum_{j=1}^8\alpha_{n,k_x,k_y}(j)\frac{1}{\sqrt{N}}\exp{i(\vec{k}+(j-1)\vec{Q})\cdot \vec{R}}
\end{equation}
So, by definition, the local number density is
\begin{equation}
n(\vec{R})=2\times{\sum_{n,k_x,k_y}}^{\prime}\left\vert{\psi_{n,k_x,k_y}(\vec{R})}\right\vert^2
\end{equation}
Here the prime  in the sum means that all {\em occupied states with energy below the chemical potential} are considered.  The factor of 2 is for spin and  the summation over $k_x,k_y$ is performed
in the RBZ.
For different parameter sets the numerical results are:
\begin{itemize}

\item{Parameter set 1\\}
$W_0=0.71t, V_c=0.05t, \mu=-0.78t, x=11.73\%$\\
Averaged number density of electron is $n=0.894$ per site, while the estimated deviation is about $\delta n =0.059 $ per site.
So $\delta n/ n=6.6\%$.

\item{Parameter set 2\\}
$W_0=0.65t, V_c=0.05t, \mu=-0.83t, x=12.46\%$\\
Averaged number density of electron is: $n =0.893$ per site, while the estimated deviation is about $\delta n = 0.062 $ per site.
So $\delta n/n=6.9\%$.
\end{itemize}
Of course, for both cases, the period of the CDW modulation is $4a$, where $a$ is the lattice spacing. It is also interesting to calculate the ratio of the 
modulation of the {\em local density of states} to the average density of states at the {\em Fermi energy}; we find $\delta\rho(\mu)/\rho(\mu)\approx 13-15\%$ depending on the parameters.
As pointed out in Ref.~\onlinecite{Yao:2011}, this leads to an estimate of the corresponding variation of the Knight shift.

\subsection{Oscillation spectra in the presence of correlated disorder}
Previously it was found from the consideration of $1/8$ magnetic antiphase stripe order that there is a remarkable variety of possible Fermi surface reconstructions depending on the choice of parameters.~\cite{Millis:2007} This is also true for the incommensurate period-8 DDW. In contrast, two-fold commensurate DDW order leads to much lesser variety. While this is more satisfying, period-4 charge modulation observed in NMR measurements~\cite{Julien:2011} and the non-existence of the  larger hole pocket commensurate with the Luttinger sum rule have forced us to take seriously the period-8 DDW. This requires a judicious choice of parameters of the model.

Although we cannot constrain the parameters uniquely, we have used a number of guiding principles.  First, disorder was chosen to be correlated with a length scale $\ell_{D}$ smaller than the transverse width of the strip, $M a$. Since the YBCO samples studied appear to have lesser degree of disorder than the intrinsic disorder of NCCO, the white noise disorder did not appear sensible. Because the experimentally measured charge modulation in NMR is $0.03\pm 0.01 e$, it is necessary to keep $V_{c}$ small enough to be consistent with experiments. A value of $V_{c}$ in the neighborhood of $0.05 t$ seemed reasonable. Of course, this could be adjusted to agree precisely with experiments, but this would not have been very meaningful.

The band structure parameter $t$ was chosen to be $0.15 eV$ as opposed to LDA value of $0.38 eV$. Although reliable ARPES measurements are not available for YBCO, measurements in other cuprates have indicated that the bandwidth is renormalized by at least a factor of 2. Had we chosen $t=0.38 eV$, the agreement with specific heat measurements would have been essentially perfect, but we could not see any justification for this. The parameters $t'/t$ and $t''/t'$ are same as the commonly used LDA values, as the shape of the Fermi surface in most cases appear to be given correctly by  LDA. We searched the remaining parameters, $\mu$, $g_{V}$ and $W_{0}$, extensively. There are a number of issues worth noting. Oscillation spectra  hardly ever show any substantial evidence of harmonics, which should be used as a constraining factor. Moreover, as we believe that it is the electron pocket that is dominant in producing negative $R_{H}$, it is necessary that we do not employ parameters that wipe out the electron pocket altogether. The coexistence of  electron and  hole pockets give a simple explanation of the oscillations of $R_{H}$ as a function of the magnetic field.  We generically found hole pocket frequencies in the range $150-300 \text{T}$. This is one of our crucial observations. It implies that to resolve clearly such a slow frequency, one must go to much higher fields than are currently possible. We argue that this may be a plausible reason why the hole pocket has not been observed except for one experiment which goes up  to $85 \text{T}$; in this experiment some evidence of a $250 \text{T}$ frequency is observed.~\cite{Singleton:2010}

\begin{figure}[htbp]
\begin{center}
\includegraphics[width=\linewidth]{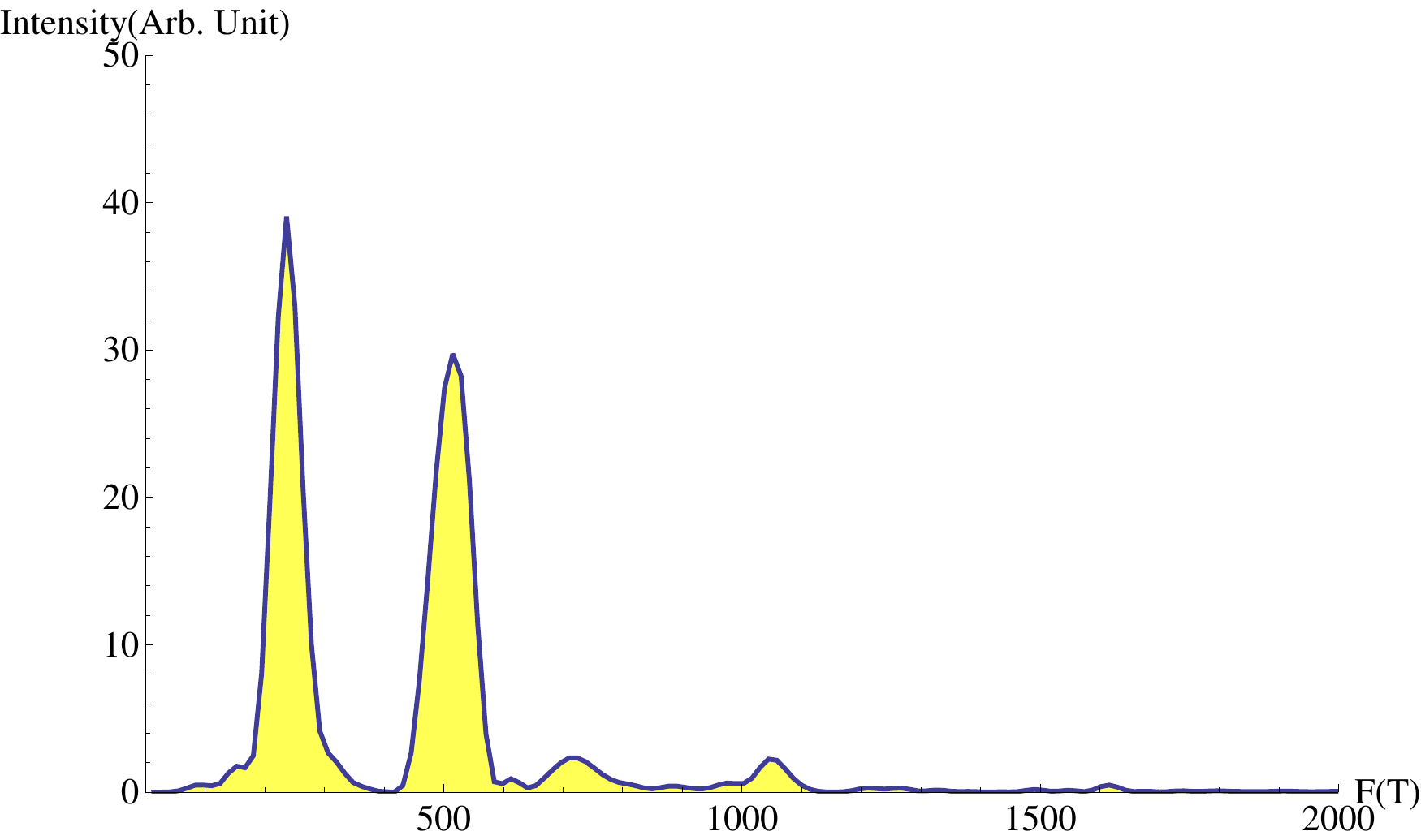}
\caption{Fourier transform of the oscillation spectra after a background subtraction with a cubic polynomial.  $W_{0}=0.71 t$, $V_{c} = 0.05t$, $\mu = -0.78t$, $M=30\;  a$,  $N=10^{5}\;  a$, $\ell_{D}=8 \; a$, $g_{V}=0.1t$. Doping is $11.73\%$.}
\label{FigB1}
\end{center}
\end{figure}

\begin{figure}[htbp]
\begin{center}
\includegraphics[width=\linewidth]{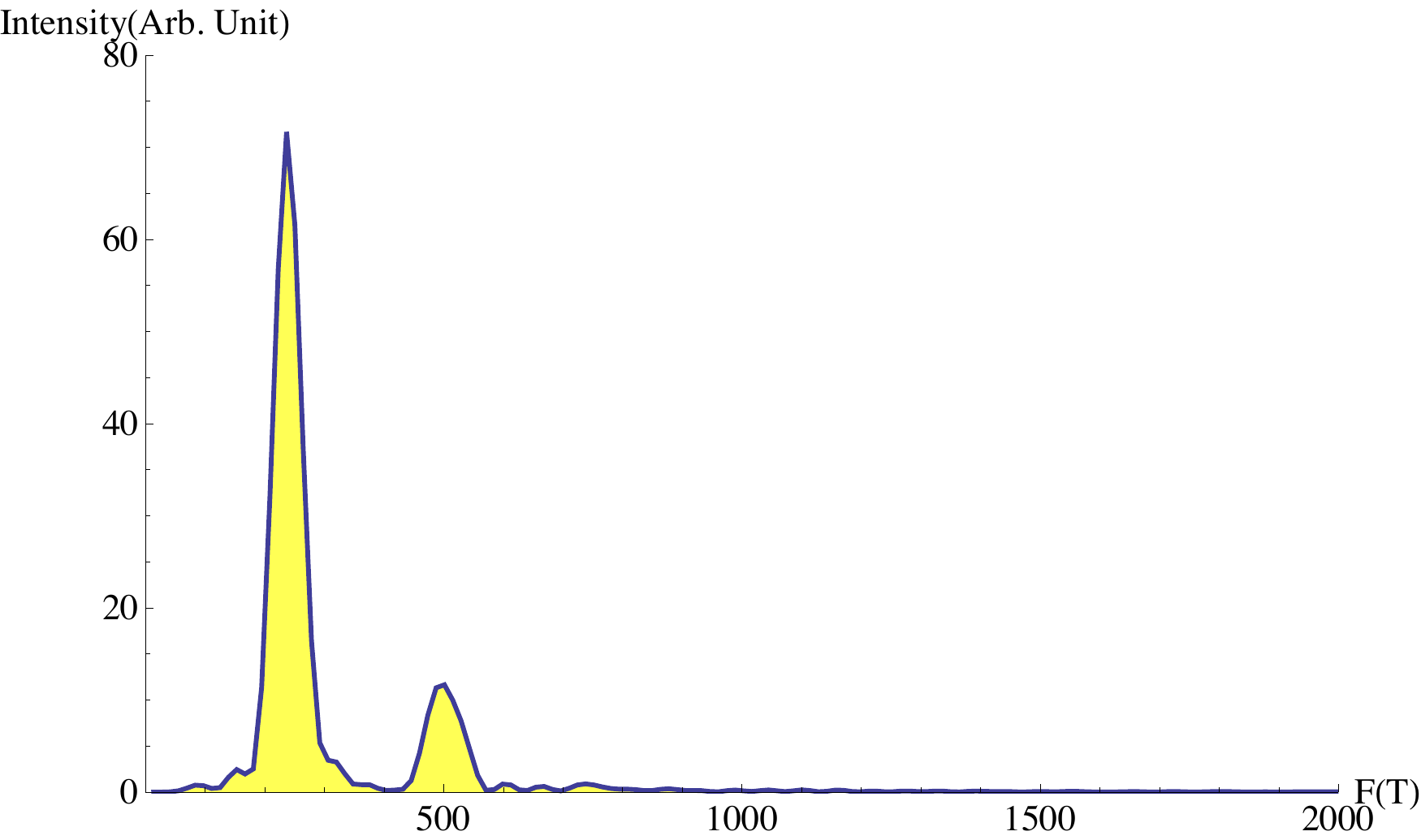}
\caption{Fourier transform of the oscillation spectra after a background subtraction with a cubic polynomial. $W_{0}=0.71 t$, $V_{c} = 0.05t$, $\mu = -0.78t$, $M=30\; a$, $N=10^{5}\;  a$, $\ell_{D}=8\; a$, $g_{V}=0.3t$. Doping is $11.73\%$.}
\label{FigB2}
\end{center}
\end{figure}

\begin{figure}[htbp]
\begin{center}
\includegraphics[width=\linewidth]{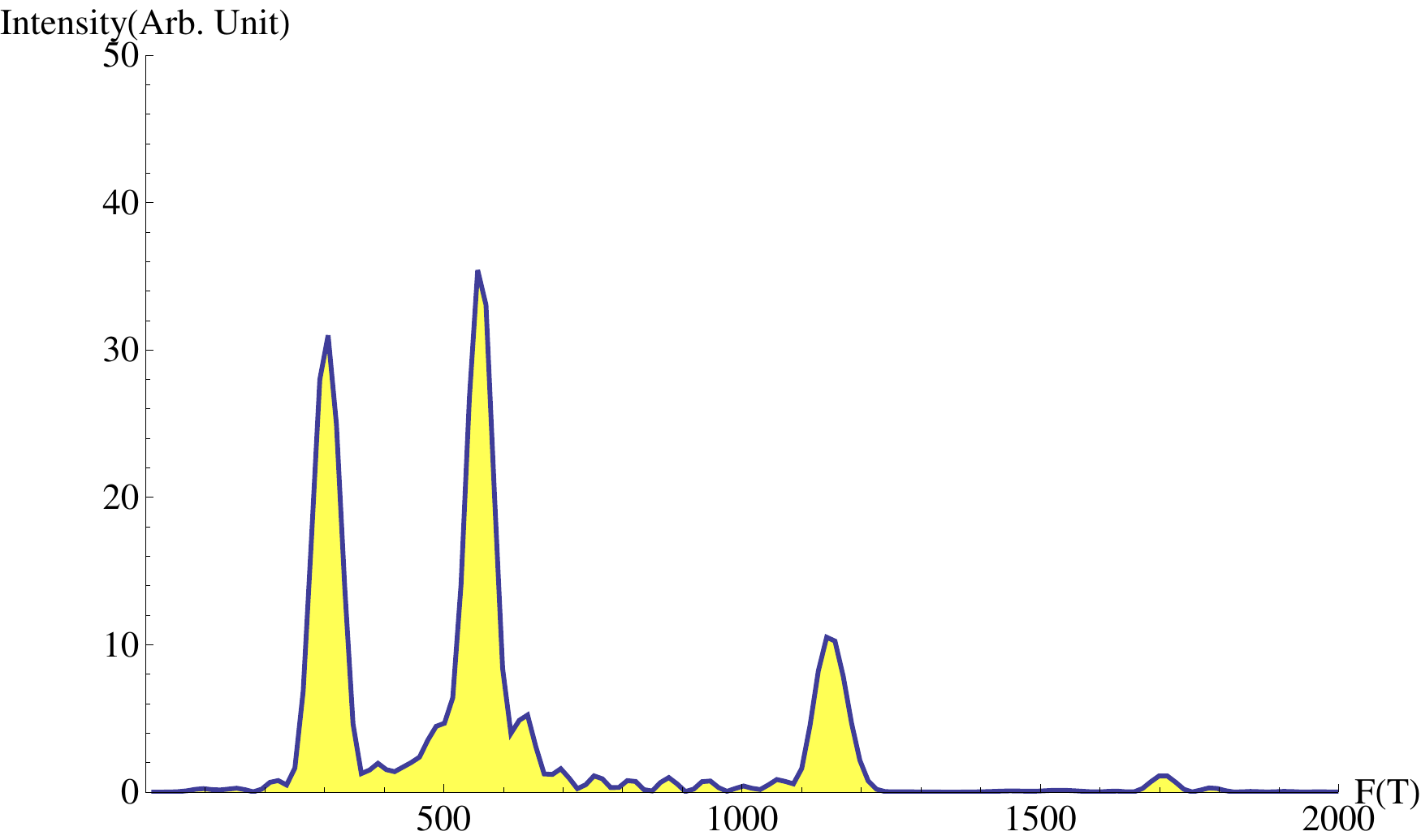}
\caption{Fourier transform of the oscillation spectra after a background subtraction with a cubic polynomial. $W_{0}=0.65 t$, $V_{c} = 0.05t$, $\mu = -0.83 t$, $M=30\; a$, $N=10^{5}\;  a$,  $\ell_{D}=8\;  a$, $g_{V}=0.1t$. Doping is $12.46\%$.} 
\label{FigC1}
\end{center}
\end{figure}

\begin{figure}[htbp]
\begin{center}
\includegraphics[width=\linewidth]{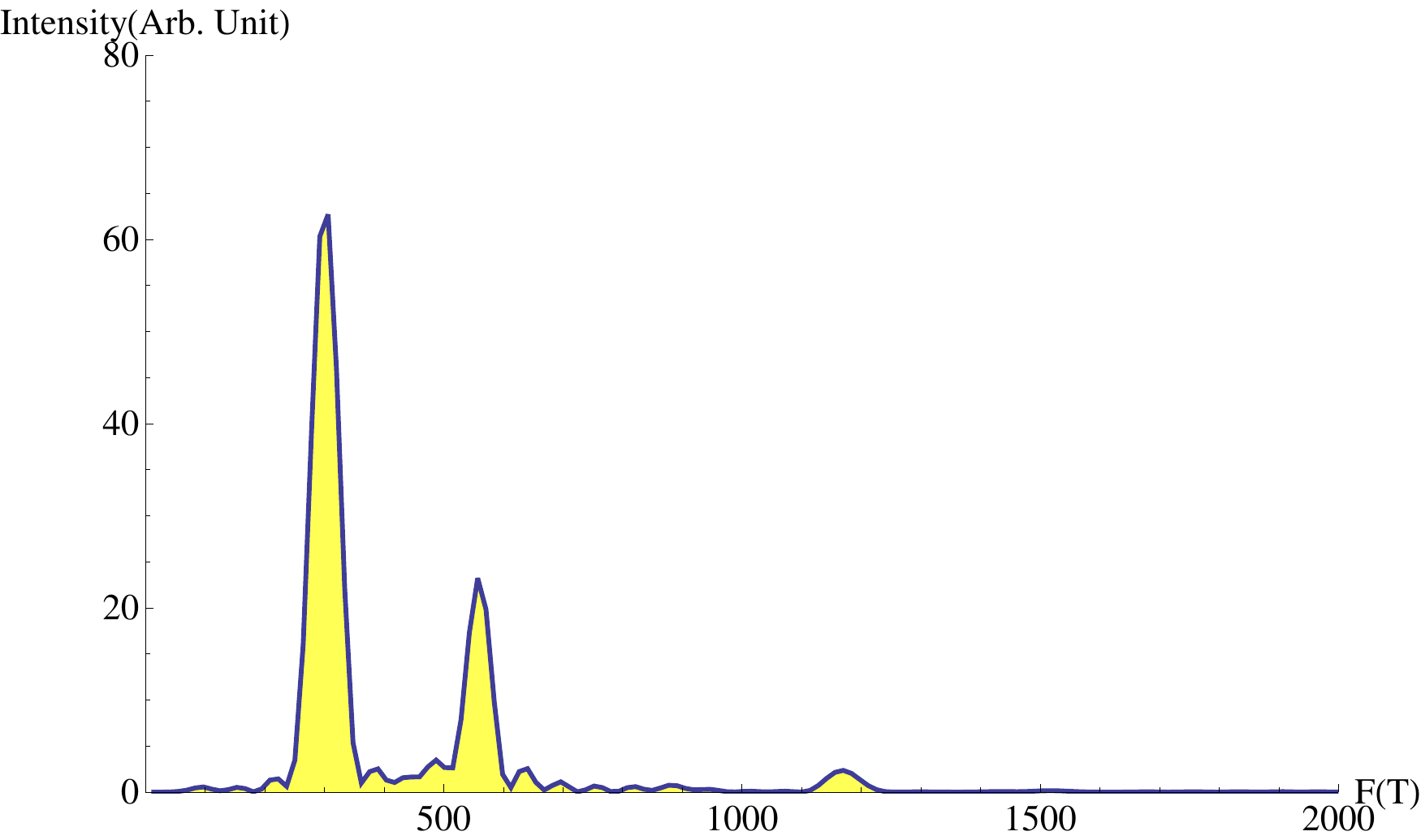}
\caption{Fourier transform of the oscillation spectra after a background subtraction with a cubic polynomial. $W_{0}=0.65 t$, $V_{c} = 0.05t$, $\mu = -0.83 t$, $M=30\; a$, $N=10^{5}\;  a$, $\ell_{D}=8\;  a$, $g_{V}=0.2t$. Doping is $12.46\%$.} 
\label{FigC2}
\end{center}
\end{figure}

A further constraining fact is that no evidence of magnetic breakdown is observed in YBCO, while in NCCO it is clearly present. This implies that our parameters should be consistent with this fact. The DDW gap and the disorder level are consistent with the observed data. Overall we find satisfactory consistency with doping levels between $11-12.5\%$ within our calculational scheme. Lower doping levels produce less satisfactory agreement, but can be made better with further adjustment of parameters, but we have avoided fine tuning as much as possible. The broad brush picture can already be seen in the oscillation spectra in Figs.~ \ref{FigB1}, ~\ref{FigB2}, ~\ref{FigC1}, and ~\ref{FigC2}. Two general trends are that electron pockets dominate at higher doping levels within the range we have checked, and an increase in disorder intensity reduces the intensity of the Fourier spectra of the electron pockets. A few harmonics are still present.

\section{Discussion and outlook}
The complex materials physics of high temperature superconductors lead to a fairly large number of dimensionless parameters. Thus, it is not possible
to frame a unique theory. Tuning these parameters can indeed lead to many different phases. However, there may be a general framework that could determine the overall picture. To be more specific, let us consider the quantum oscillation measurements that we have been discussing here.
\begin{itemize}
\item Are the applied magnetic fields sufficiently large to essentially destroy all traces of superconductivity and thereby reveal the underlying normal state from which superconductivity develops? While
for NCCO this is clearcut because $H_{c2}$ is less than $10 \text{T}$, while the quantum oscillation measurements are carried out between $30 - 65 \text{T}$, far above $H_{c2} $. For YBCO lingering doubts remain. However,   one may argue that the high field measurements
are such that one may be in a vortex liquid state where  the slower vortex degrees of freedom may simply  act as quenched
disorder to the nimble electrons. This is the picture we have adopted here.
\item The emergent picture of Fermi pockets are seemingly at odds with ARPES, unless only half the pocket is visible in ARPES, as was previously argued.~\cite{Chakravarty:2003} On the other hand, reliable ARPES in YBCO is not
available. For electron doped NCCO or PCCO this appears not be true.~\cite{Armitage:2009}
\item Almost all scenarios place the observed  electron pockets at the anti-nodal points in the Brillouin zone, while many other experiments would require the pseudogap to be maximum there.
One may, however,  question if there is only one pseudo gap.
\item Quantum oscillations of $R_{H}$ are easier to explain if there are  at least two closed pockets in the Boltzmann picture.~\cite{Chak2:2008} Thus associated with the electron pocket there must be 
a hole pocket or vice versa. This is not a problem with NCCO, as we have shown how magnetic breakdown,~\cite{Eun:2010} and a greater degree of intrinsic disorder, provides a simple resolution as to  why only one pocket, in this case a small but prominent hole pocket is seen. In any case, oscillations of $R_{H}$ in NCCO is yet to be measured. With respect to YBCO this becomes a serious problem. Any commensurate picture would lead to a hole pocket of frequency about twice that of the electron pocket frequency if the Luttinger sum rule is to be satisfied. Despite  motivated effort no evidence in this regard has emerged. An escape from the dilemma is to propose an incommensurate picture in which the relevant electron pocket is accompanied by a much {\em smaller} hole pocket and  some open orbits, as we have done here. In order to convincingly observe such small hole pocket, one would require extending these measurements  to almost impossibly higher fields; see, however, Ref.~\onlinecite{Singleton:2010}.
\item All oscillation measurements to date have been convincingly interpreted in terms of the Lifshitz-Kosevich theory for which the validity of Fermi liquid theory and the associated Landau levels seem to be obligatory. Why should the normal state of an under doped cuprate behave like a Fermi liquid?
\item The  contrast between  electron and hole doped cuprates is interesting. In NCCO the crystal  structure consists of a  single CuO plane per unit cell, and, in contrast to YBCO, there are no complicating
 chains, bilayers, ortho-II potential, stripes, etc.~\cite{Armitage:2009} Thus, it would appear to be ideal for gleaning the mechanism of quantum oscillations. On the other hand, disorder in NCCO  is significant. It is believed that well-ordered chain materials of  YBCO contain much less disorder by comparison.  
 \item In YBCO, studies involving tilted field seem to rule out  triplet order parameter, hence SDW.~\cite{Ramshaw:2011,Sebastian:2011} Moreover, from NMR measurements at high fields, there appears to be no evidence of a 
static spin density wave order in YBCO.~\cite{Julien:2011} Similarly there is no evidence of SDW order in fields as high as $23.2T$ in $\mathrm{YBa_{2}Cu_{4}O_{8}}$~\cite{Zheng:1999}, while quantum oscillations are clearly observed in this material.~\cite{Yelland:2008,Bangura:2008}  Also no such evidence of SDW  is found up to $44T$ in $\mathrm{Bi_{2}Sr_{2-x} La_{x}CuO_{6+\delta}}$.~\cite{Kawasaki:2011} At present, results from high field NMR  in NCCO does not exist, but measurements are  in progress.~\cite{Brown:2011} The zero field neutron scattering measurements  indicate very small spin-spin correlation length in
the relevant doping regime.~\cite{Motoyama:2007} Energetically a  perturbation even as large as  $45 T$ field is weak.~\cite{Nguyen:2002}

\item As to singlet order, relevant to quantum oscillations,~\cite{Garcia-Aldea:2011,*Norman:2011,*Ramazashvili:2011}  charge density wave  is a possibility, which has recently found some support in the 
high field NMR measurements  in YBCO.~\cite{Julien:2011} But since the mechanism is helped by the oxygen chains, it is unlikely that the corresponding NMR measurements in NCCO will find such a charge order. Moreover, the observed charge order in YBCO sets in at a much lower temperature ($ 20- 50 K$) compared to the pseudogap. Thus the charge order may be parasitic.
As to  singlet DDW, there are two neutron scattering measurements that seem to provide evidence for it.~\cite{Mook:2002,*Mook:2004} However, these measurements have not been confirmed by further independent experiments. However, DDW order should be considerably hidden in NMR involving nuclei at high symmetry points, because the orbital currents should cancel.
\end{itemize}
As mentioned above, a  mysterious feature of quantum oscillations in YBCO is the fact that only one type of Fermi pockets is observed. If two-fold commensurate density wave is 
the mechanism, this will violate the Luttinger sum rule.~\cite{Luttinger:1960,*Chubukov:1997,*Altshuler:1998,Chak2:2008} We had previously provided an explanation of this phenomenon in terms of disorder arising from both
defects and vortex scattering in the vortex liquid phase;~\cite{Jia:2009} however, the arguments are not unassailable. In contrast, for NCCO, the experimental results are 
quite consistent with a  simple theory presented previously. The present work, based on incommensurate DDW, may provide another, if not a more plausible alternative in YBCO.

The basic question as to  why Fermi liquid concepts should apply remains an important  unsolved mystery.~\cite{Chakravarty:2011} It is possible that if the state
revealed by applying a high magnetic field has a broken symmetry with an order parameter (hence a gap), the low energy excitations will be quasiparticle-like, not a spectra with a  
branch cut, as in variously proposed strange metal phases. 

\section{Acknowledgments}
We would like to thank S. Kivelson and B. Ramshaw for very helpful discussion. This work is supported by NSF under the Grant DMR-1004520.

\end{document}